\documentstyle[12pt]{article}
\begin{document}
\newcommand{\p}[1]{(\ref{#1})}
\newcommand{\be}{\begin{equation}}
\newcommand{\ee}{\end{equation}}
\newcommand{\nn}{\nonumber}
\newcommand{\sect}[1]{\setcounter{equation}{0}\section{#1}}
\renewcommand{\theequation}{\arabic{equation}}
\newcommand{\vs}[1]{\rule[- #1 mm]{0mm}{#1 mm}}
\newcommand{\hs}[1]{\hspace{#1mm}}
\newcommand{\mb}[1]{\hs{5}\mbox{#1}\hs{5}}
\newcommand{\Db}{{\overline D}}
\newcommand{\DDB}{\left[ D,\Db \right]}
\newcommand{\bea}{\begin{eqnarray}}
\newcommand{\eea}{\end{eqnarray}}
\newcommand{\PL}[1]{Phys.\ Lett.\ {\bf #1}}

\begin{titlepage}
\begin{flushright}
JINR E2-96-344\\
hep-th/9609191\\
September 1996
\end{flushright}
\vskip 1.0truecm
\begin{center}
{\large \bf NEW INTEGRABLE EXTENSIONS OF N=2 KdV AND BOUSSINESQ  
HIERARCHIES}
\vglue 2cm
{\bf E. Ivanov and S. Krivonos}
\vglue 1cm
{\it Bogoliubov Laboratory of Theoretical Physics,
JINR,\\
141 980 Dubna, Moscow Region, Russia}
\end{center}

\vspace{2cm}

\begin{abstract}
We construct a new variety of $N=2$ supersymmetric integrable systems 
by junction of pseudo-differential superspace Lax operators for 
$a=4$, $N=2$ KdV and multi-component $N=2$ NLS hierarchies. As 
an important particular case, we obtain Lax operator for $N=4$ super 
KdV system. A similar extension of one of $N=2$ super Boussinesq 
hierarchies is given. We also present a minimal $N=4$ supersymmetric 
extension of the second flow of $N=4$ KdV hierarchy and comment 
on its possible integrability. 
\end{abstract}

\end{titlepage}

\noindent{\bf 1. Introduction.} For the last years, 
$N=2$ supersymmetric 
hierarchies of integrable equations (of the KP, KdV and NLS types) 
attracted an increasing interest, 
mainly due to their potential physical applications in 
non-perturbative $2D$ supergravity, superextensions of matrix models 
and topological field theories (see, e.g. [1]). Unexpected 
interrelations between these hierarchies were revealed and 
different manifestly $N=2$ supersymmetric Lax representations for 
them were constructed [2-5]. 

In ref. \cite{KST} a new form of the Lax representation 
for the $a=4$, $N=2$ KdV hierarchy through the pseudo-differential Lax 
operator was proposed\footnote{We use the following conventions about 
the algebra of $N=2$, $1D$ superspace derivatives 
$$
D=\frac{\partial}{\partial\theta}-
 \frac{1}{2}\bar\theta\frac{\partial}{\partial z},\quad 
\Db=\frac{\partial}{\partial\bar\theta}-
 \frac{1}{2}\theta\frac{\partial}{\partial z},\;\;\;
\left\{ D,\Db \right\} = -\frac{\partial}{\partial z},\quad
\left\{ D,D \right\}=\left\{ \Db,\Db \right\} =0\;, 
$$
$$ 
(D)^{\dagger} = -\Db\;, 
(\frac{\partial}{\partial z})^{\dagger} = 
- \frac{\partial}{\partial z}\;. 
$$}  
\be 
L=\partial -2J-2\Db\partial^{-1}(DJ) \quad , \label{lax}
\ee
\be
\frac{\partial}{\partial t_k} L=\left[ (L^k)_{\geq1},L\right]\;.
\label{laxeq}
\ee
Here, $J=J(z,\theta, \bar \theta)$ is a general $N=2$ superfield, 
$k=1,2,\ldots$ and the subscript $\geq 1$ means restriction 
to the purely differential part of $L^k$.
 
As distinct from the previously known representation with 
the differential Lax operator \cite{LM}, 
this form produces the whole set of bosonic conserved quantities 
for the $a=4$, $N=4$ KdV (both of odd and even scale dimensions) 
by the general formula 
\be \label{ham}
H^n = \int dZ \left( L^n \right)_0 \quad .
\ee
Note a non-standard definition of the residue of the powers of the Lax 
operator in \p{ham}. In the conventional Lax representation \cite{LM}, 
the standard definition of the residue as a coefficient before 
$[D,\bar D]\partial^{-1}$ is used, however, only odd-dimension conserved 
charges can be directly constructed within its framework. 

The Lax representation \p{lax}, \p{laxeq} is also advantageous in that it 
provides a link with the $N=2$ NLS hierarchy. After 
the Miura type transformation \cite{KS}
\be
J = -\frac{1}{2}F{\bar F}-\frac{1}{2}\frac{D {\bar F}'}{D {\bar F}} , 
\ee
where $F, \bar F$ are chiral and anti-chiral $N=2$ superfields 
($DF = \Db \bar F = 0$), 
some further similarity transformation of the Lax operator and 
passing to its adjoint, one gets new Lax operator \cite{KST}
\be  \label{lax2}
L'= \partial - F{\bar F}-F\Db\partial^{-1}
     (D{\bar F}). \label{laxff}
\ee
It gives rise, via the representation \p{laxeq}, to the minimal 
$N=2$ extension of NLS hierarchy. In ref. \cite{BKS} 
multi-component generalizations of the Lax operator \p{laxff} 
were constructed. 

The above correspondence between two Lax operators generalizes 
the situation known in the purely bosonic case for the $R-S$ system. 
The bosonic analogs of the $N=2$ Lax operators \p{lax}, \p{laxff} 
are as follows \cite{{BX},{FK}}
\be  \label{laxb1}
L^{(1)} = \partial + R\frac{1}{\partial -S}\;, 
\ee
\be \label{laxb2}
L^{(2)} = \partial + r \partial^{-1} {\bar r}\;.
\ee
They are related to each other by a generalized 
Miura transformation
\be\label{miura1}
R=r{\bar r}, \quad S=\frac{{\bar r}'}{\bar r} \; .
\ee
One may construct a "hybrid" Lax operator as the sum 
$L^{(1)} + L^{(2)}$. However, the latter does not produce a new 
system as it can be reduced to a two-component generalization 
of \p{laxb2} by the transformation \p{miura1} (a self-consistent 
generalization of \p{laxb2} is to attach an index $i$ 
to the field $r$ and to sum up over $i$).

Since the $N=2$ Lax operators \p{lax}, \p{lax2} are related in a rather 
obscure way (through an additional similarity transformation 
and conjugation), an analogous hybridization procedure in this 
case might yield new integrable systems, more general than the 
$N=2$ KdV or $N=2$ NLS hierarchies. The basic aim of the present 
note is to demonstrate that this is indeed the case. We also elaborate 
on some generalizations and consequences of this fact.
\vspace{0.3cm}

\noindent{\bf 2. Hybrid $N=2$ KdV-NLS hierarchies.} Let us introduce $M$ 
pairs of chiral and anti-chiral
dimension $1/2$ superfields $F^i,{\bar F}^i$. They can be either fermionic 
or bosonic and in general are not obliged to be conjugated to each other 
(in this case $J$ should also be complex). 
The $U(1)$ charges of $F$ and $\bar{F}$ inside each pair are taken to 
be opposite, but the relative charges of different pairs are noway fixed. 
We construct the following Lax operator
\be
L_1=\partial -2J-2\Db\partial^{-1}(DJ) -
\sum_i F^i\Db\partial^{-1}(D{\bar F}^i) +
\sum_i \Db\partial^{-1}(D(F^i{\bar F}^i)) \quad . \label{lax1}
\ee
We have checked that it gives rise, through the same 
Lax equation \p{laxeq}, to the self-consistent hierarchy of the 
evolution equations. 
When $F^i = {\bar F}^i = 0$, the operator $L_1$ is 
reduced to \p{lax}. With 
$J = \frac{1}{2} \sum_i F^i{\bar F}^i $, the multi-component 
generalization of the Lax operator 
\p{lax2} is recovered \cite{BKS}. Respectively, the related hierarchy 
is reduced to either $a=4$, $N=2$ KdV or multi-component 
NLS ones. Thus we have got new integrable extensions of both 
these hierarchies. 

Explicitly, the second and third flows are as follows
\bea
\frac{\partial J}{\partial t_2} & = &-\left[ D,\Db \right] J'-4JJ'+
 \sum_i (\Db F^i D{\bar F}^i)' \quad , \nn \\
\frac{\partial F^i}{\partial t_2} & = &{F^i}{}''+4D(J\Db F^i) 
\quad ,\nn \\
\frac{\partial {\bar F}^i}{\partial t_2} & = & -{{\bar F}^i}{}''+
      4\Db (JD{\bar F}^i)\;, 
\label{2flow1}
\eea
\bea \label{3flow1}
\frac{\partial J}{\partial t_3} & =  & J'''+3\left( \DDB J J \right)'+
\frac{3}{2}\left( \DDB J^2\right)'+4\left( J^3\right)'-
\frac{3}{2} \sum_i \left(  F^i{}'{\bar F}^i{}'+
 4 J\Db F^i D{\bar F}^i\right)'\;, \nn \\
\frac{\partial F^i}{\partial t_3} & =  & F^i{}'''+
D \left( 6 \Db (JF^i{}')-12 J^2\Db F^i +
         3 \sum_j D{\bar F}^j\Db F^j \Db F^i \right)\;, \nn \\
\frac{\partial {\bar F}^i}{\partial t_3} & =  & {\bar F}^i{}'''-
\Db \left( 6 D (J{\bar F}^i{}')+12 J^2 D {\bar F}^i -
 3 \sum_j D{\bar F}^j\Db F^j D {\bar F}^i \right)\;.
\eea
An interesting peculiarity of eqs. \p{2flow1}, \p{3flow1} is that 
the dimension 1/2 
superfields $F^i$ and ${\bar F}^i$ appear in the nonlinear terms 
only under 
spinor derivatives, i.e. as $\Phi^i = \Db F^i$,  
$\bar \Phi^i = D{\bar F}^i$, 
$\Db \Phi^i = D \bar \Phi^i = 0$ (for instance, 
$F^{i}{}'$ = $- D\bar{\Phi}^i$, etc). 
Acting on both sides of the $F$ equations by $D$, $\Db$, one can rewrite 
the above sets entirely in terms of the chiral and anti-chiral dimension 
1 superfields $\Phi^i$, $\bar \Phi^i$. In this sense $F^i$ and 
$\bar F^i$ can be regarded as prepotentials of the superfields 
$\Phi^i$, $\bar \Phi^i$ in some fixed gauge with respect to the 
prepotential gauge freedom. Note that the relation between the 
superfields $\Phi^i$ and $F^i$ is invertible 
\be \label{invrel}
F^i = - D\;\partial^{-1} \Phi^i\;,\;\;  \bar{F}^i = - \Db \;\partial^{-1} 
\bar{\Phi}^i \;.
\ee

It is somewhat surprising 
that for one pair of mutually conjugated fermionic superfields 
$F, {\bar F} = F^{\dagger}$ and real $J$, the systems \p{2flow1},
\p{3flow1}, being rewritten in terms of the superfields 
$\Phi, \bar{\Phi}$, coincide with the second and third flows 
of the $N=4$, $SU(2)$ KdV hierarchy constructed in \cite{{DI},{DIK}} 
(actually, for one of possible equivalent choices of the $SU(2)$ 
breaking parameters in it, $a=4, \;b=0$). For instance, 
the second flow equations take the form 
\bea  \label{n42}
\frac{\partial J}{\partial t_2} & = & - \left[ D,\Db \right] J'-4JJ' + 
(\Phi \bar{\Phi})' \quad , \nn \\
\frac{\partial \Phi}{\partial t_2} & = & \Phi {}''+4\Db D \;(J\Phi) 
\quad , \nn \\
\frac{\partial \bar{\Phi}}{\partial t_2} & = & - \bar{\Phi}{}''+
      4D \Db\; (J \bar{\Phi}) \quad .
\eea
It is easy to check the covariance of this set under the transformations 
of an extra hidden $N=2$ supersymmetry \cite{DIK}
\be  \label{n4susy}
\delta J = {1\over 2} \epsilon D \Phi + 
{1\over 2} \bar{\epsilon} \Db \bar{\Phi}\;, \;\;
\delta \Phi = -2 \bar{\epsilon} \Db J\;, \;\; 
\delta \bar{\Phi} = -2 \epsilon DJ\;.
\ee
Here, $\epsilon, \bar{\epsilon}$ are mutually conjugated 
Grassmann parameters. Together with the explicit $N=2$ 
supersymmetry these transformations constitute $N=4$ supersymmetry 
in one dimension. Note that an equivalent realization in terms of 
the superfields $F, \bar{F}$ is non-local  
\be  \label{n4susyF}
\delta J = - {1\over 2} \epsilon F{}'  
- {1\over 2} \bar{\epsilon} \bar{F}{}'\;, \;\; 
\delta F = -2 \bar{\epsilon} D\partial^{-1} \Db J\;, \;\; 
\delta \bar{F} = -2 \epsilon \Db \partial^{-1} DJ\;.
\ee

Thus in this particular case the Lax equation \p{laxeq} solves 
the problem of constructing the Lax representation for $N=4$ KdV 
hierarchy. Actually, 
this proves the very existence of such an 
integrable hierarchy, the fact conjectured in \cite{{DI},{DIK}} 
on the ground of the existence of higher order non-trivial conservation 
laws and the bi-hamiltonian property for this system. Now it is a matter 
of straightforward computation, using the general formula \p{ham}, 
to reproduce the conserved quantities which were constructed 
in \cite{DIK} by the "brute force" method.

Note that another choice of the relation between $F$ and $\bar{F}$, 
\be
\bar{F} = -F^{\dagger}\;, \; (\;\bar{\Phi} = - \Phi^{\dagger}\;)
\ee 
(with keeping $J$ real as before) results in a different system. 
It formally coincides with \p{n42} but 
is invariant under the following modification of the transformations 
\p{n4susy1} 
\be  \label{n4susy1}
\delta J = {1\over 2} \epsilon D \Phi - 
{1\over 2} \bar{\epsilon} \Db \bar{\Phi}\;, \;\;
\delta \Phi = 2 \bar{\epsilon} \Db J\;, \;\; 
\delta \bar{\Phi} = -2 \epsilon DJ\;.
\ee
Together with the manifest $N=2$ supersymmetry these constitute 
a ``twisted'' $N=4$ supersymmetry (commutator of two such 
transformations yields $\frac{\partial}{\partial z}$ with the 
opposite sign as compared to the manifest $N=2$ 
supersymmetry transformations and the transformations 
\p{n4susy} or \p{n4susyF}).

For a greater number of pairs $F, \bar{F}$ we get the extensions which, 
to our knowledge, were not considered before. The bosonic sector of 
the generic system \p{2flow1} reads 
\bea
\frac{\partial J}{\partial t_2} & = &
      -T'-4JJ'-\sum_i ({\bar H}^iH^i)'  \quad , \nn \\
\frac{\partial T}{\partial t_2} & = &-J'''+
       \sum_i ({\bar H}^i{H^i}'-{{\bar H}^i}{}'H^i)'-4(TJ)' 
\quad , \nn \\
\frac{\partial H^i}{\partial t_2} & = &-{H^i}{}''+2TH^i-4J{H^i}{}'-
      2J'H^i\quad ,  \nn \\
\frac{\partial {\bar H}^i}{\partial t_2} & = &{{\bar H}^i}{}''-
   2T{\bar H}^i-4J{{\bar H}^i}{}'-2J'{\bar H}^i \quad . \label{bossec1}
\eea
Here the bosonic components are defined as
\be
J=J|,T=\DDB J|,H=-D{\bar F}|, {\bar H}=\Db F|
\ee
and $|$ means the restriction to the $\theta=\bar\theta=0$ 
parts.
\vspace{0.3cm}

\noindent{\bf 3. An extension of $N=2$ Boussinesq hierarchy.} One may 
wonder whether similar extensions are possible for generalized $N=2$ 
KdV hierarchies associated with $N=2$ $W_n$ algebras as the second 
hamiltonian structures. In refs. \cite{{BIKP},{P}}  
$N=2$ superfield differential Lax operator for the $\alpha=-1/2$, $N=2$ 
Boussinesq equation (with $N=2$ $W_3$ the second hamiltonian structure) 
was constructed
\be\label{BL1}
L=D\left( \partial^2-3J\partial-T-\frac{3}{2}J'-\frac{1}{2}\DDB J+
  2J^2\right)\Db \;.
\ee
In trying to modify it along the above lines we have found that the only 
consistent modification is the following one  
\be\label{BL2}
L_1=D\left( \partial^2-3J\partial-T-\frac{3}{2}J'-\frac{1}{2}\DDB J+
  2J^2+ \sum_i \Psi^i\partial^{-1}{\bar\Psi}^i\right)\Db
\ee
where M pairs of fermionic (bosonic) chiral and anti-chiral 
superfields $\Psi^i,{\bar \Psi}^i$ with dimension $3/2$ were introduced. 
This new Lax operator \p{BL2} gives rise through the Lax equation
\be
\frac{\partial L_1}{\partial t_2} =
    \left[ \left( L_1^{\frac{2}{3}}\right)_{>1},L_1 \right]
\ee
to the following extension of $N=2$ Boussinesq hierarchy
\bea \label{SB}
\frac{\partial J}{\partial t_2} & = & 2T'+\DDB J'-2JJ'\;, \nn \\
\frac{\partial T}{\partial t_2} & = & -2J'''+10\left( \Db J D J\right)'+
 4J'\DDB J+2 J\DDB J'+4J'J^2 +6\Db J DT+ \nn \\
 & & 6D J\Db T+6J'T+2 JT'-6\sum_i \left( \Psi^i{\bar\Psi}^i\right)' \;,
\nn \\
\frac{\partial \Psi^i}{\partial t_2} & = & 3\Psi^i{''}-6J\Psi^i{}'+
         6D J\Db \Psi^i \;, \nn \\
\frac{\partial {\bar\Psi}^i}{\partial t_2} & = & -3{\bar\Psi}^i{''}-
    6J{\bar\Psi}^i{}'+
         6\Db JD {\bar\Psi}^i\;. 
\eea
We checked the existence of first non-trivial higher-order conservation 
laws for this hierarchy by using the general formula 
\be 
H_k = \int dX \mbox{Res} L^{k/3}
\ee
(in this case, just as in the $\alpha = -1/2$, $N=2$ Boussinesq  
limit $\Psi^i = \bar{\Psi}^i = 0$, the residue of $N=2$ 
pseudo-differential operators is defined in the standard way 
as the coefficient before 
$[D, \Db] \partial^{-1}$). We found that there exist conserved 
quantities $H_k$ of {\it all} scale dimensions $k$, 
while in the pure $\alpha = -1/2$, $N=2$ Boussinesq case $H_{3n}$  
drop out \cite{BIKP}. The evident reason for non-existence 
of $H_{3n}$ in this case and their presence in the modified 
case is that the $N=2$ Boussinesq Lax operator \p{BL1} is differential 
and so is its any integer power, while \p{BL2} is pseudo-differential. 
One can check by explicit computation that the densities 
${\cal H}_k, \;k = 3n$ vanish when $\Psi^i, \bar{\Psi}^i$ are 
equated to zero.

It is an open question whether one can consistently reduce the above 
hierarchy to the form containing only extra superfields 
$\Psi^i, \;\bar\Psi^i$, similarly to the case of extended $N=2$ KdV 
hierarchy discussed in the previous section (recall 
that it is achieved by putting $J = \frac{1}{2} \sum_i F^i{\bar F}^i$ 
in eqs. \p{2flow1}, \p{3flow1}). 
An essential difference of the set \p{SB} from \p{2flow1}, 
\p{3flow1} consists in that in the former 
case it is impossible to trade $\Psi^i, \;\bar\Psi^i$ for their 
spinor derivatives by applying the latter to both sides of the $\Psi$ 
equations (actually, these superfields are present on their own in 
the equation for $T$ as well).
\vspace{0.3cm}

\noindent{\bf 4. A new $N=4$ supersymmetric system.} As the last 
remark, let us present a minimal $N=4$ supersymmetry preserving 
extension of the second $N=4$ KdV flow \p{n42}. 

The simplest possibility to extend the set $J, F, \bar{F}$ to some 
reducible $N=4$ supermultiplet is to add two extra pairs of 
mutually conjugated superfields $F_i, \bar{F}_i,\;i=1,2$ (for 
definiteness, we choose them fermionic) with the following 
transformation law under 
the hidden $N=2$ supersymmetry   
\be  \label{n4F12}
\delta F_1 = -\bar{\epsilon} D \bar{F}_2\;,\; 
\delta F_2 = \bar{\epsilon} D \bar{F}_1\;, \;
\delta \bar{F}_1 = - \epsilon \Db F_2\;,\; 
\delta \bar{F}_2 = \epsilon \Db F_1\;,\; 
\ee
or, in terms of $\Phi_i = \Db F_i, \bar{\Phi}_i=D{\bar F}_i$, 
\be  \label{n4fi12}
\delta \Phi_1 = \bar{\epsilon} \Db \bar{\Phi}_2\;,\; 
\delta \Phi_2 = -\bar{\epsilon} \Db \bar{\Phi}_1\;, \;
\delta \bar{\Phi}_1  = \epsilon D \Phi_2\;,\; 
\delta \bar{\Phi}_2 = - \epsilon D \Phi_1\;. 
\ee  
It is easy to check that the Lie bracket of these transformations is 
the same as for \p{n4susy}, \p{n4susyF}.

The second flow system \p{2flow1} with three extra pairs of the 
$F$ superfields, as it stands, does not respect 
covariance under \p{n4susyF}, \p{n4F12}. 
However, let us consider the following modification of it (in the 
notation through $\Phi, \bar{\Phi}$, $\Phi_i, \bar{\Phi}_i$)
\bea  \label{n42mod}
\frac{\partial J}{\partial t_2} & = & - \left[ D,\Db \right] J'-4JJ' + 
(\Phi \bar{\Phi})' - (\Phi_1 \bar{\Phi}_1)'  + (\Phi_2 \bar{\Phi}_2)'
\quad , \nn \\
\frac{\partial \Phi}{\partial t_2} & = & \Phi {}''+4\Db D \;(J\Phi 
+ {1\over 2} \Phi_1\Phi_2) \quad , \nn \\
\frac{\partial \bar{\Phi}}{\partial t_2} & = & - \bar{\Phi}{}''+
      4D \Db\; (J \bar{\Phi} + {1\over 2} \bar{\Phi}_1 \bar{\Phi}_2) 
\quad , \nn \\
\frac{\partial \Phi_1}{\partial t_2} & = & \Phi_1 {}''+ 4\beta \Db D \;
(J\Phi_1 - {1\over 2} \Phi \bar{\Phi}_2) \quad , \nn \\
\frac{\partial \bar{\Phi}_1}{\partial t_2} & = & - \bar{\Phi}_1{}''+
      4\beta D \Db\; (J \bar{\Phi}_1 - {1\over 2} \bar{\Phi} \Phi_2) 
\quad , \nn \\
\frac{\partial \Phi_2}{\partial t_2} & = & - \Phi_2 {}''+ 4\beta \Db D \;
(J\Phi_2 + {1\over 2} \Phi \bar{\Phi}_1) \quad , \nn \\
\frac{\partial \bar{\Phi}_2}{\partial t_2} & = & \bar{\Phi}_2{}''+
      4\beta D \Db\; (J \bar{\Phi}_2 + {1\over 2} \bar{\Phi} \Phi_1) 
\quad , 
\eea
$\beta$ being a parameter. One immediately checks that the modified 
system is $N=4$ supercovariant. Note that the extra superfields 
$\Phi_i$ have the dimension 1, just as $\Phi$, but their 
$U(1)$ charges are twice as smaller compared to the $U(1)$ charge of 
$\Phi$.

For the time being, we did not succeed in finding a modification 
of the Lax operator \p{lax1} which would lead to \p{n42mod} via 
the equation like \p{laxeq}. Instead we studied 
the issue of existence of higher-order non-trivial conserved quantities 
for \p{n42mod}, as the standard test for integrability. Using 
the undetermined coefficients method and the Mathematica package 
for $N=2$ superfield computations \cite{KT},  we found that at least two 
non-trivial conserved charges exist for this system 
\bea
H_2 &=& \int dX \left\{ J^2 - {1\over 2} \Phi \bar{\Phi} +{1\over 2\beta} 
\left( \Phi_1\bar{\Phi}_1 - \Phi_2\bar{\Phi}_2 \right) \right\}\;, \nn \\
H_3 &=& \int dX \left\{ {2\over 3} J^3 + \Db J DJ + {1\over 4} \Phi {}' 
\bar{\Phi} - {1\over 4\beta} \left( \Phi_1 {}' \bar{\Phi}_1  + 
\Phi_2 {}' \bar{\Phi}_2 \right) 
-J\Phi \bar{\Phi}  \right. \nn \\
&& \left.+ 
J\Phi_1 \bar{\Phi}_1 - J\Phi_2 \bar{\Phi}_2 
- {1\over 2} \Phi \bar{\Phi}_1\bar{\Phi}_2 
- {1\over 2} \bar{\Phi} \Phi_1 \Phi_2 \right\}\;. 
\eea
These quantities respect rigid $N=4$ supersymmetry and, after setting 
$\Phi_i = \bar{\Phi}_i = 0$ and further $\Phi = \Phi = 0$, are reduced 
to the same dimension conserved charges of the $N=4$, $SU(2)$ KdV and 
$a=4$, $N=2$ KdV, respectively.

We also analyzed the existence of the next conserved hamiltonian, 
$H_4$. We found that no such quantity exists, provided the relevant 
density is local 
in the superfields $\Phi_i, \bar{\Phi}_i$. Recall, however, 
that the basic objects we started with are the dimension $1/2$ 
fermionic superfields $F_i, \bar F_i$. We conjecture that the 
candidate higher-order 
conserved quantities, beginning with $H_4$, should include terms where 
these basic superfields appear on their own, with no spinor derivatives 
on them. These terms are {\it nonlocal} when written in terms of 
$\Phi_i$, $\bar{\Phi}_i$. Another possibility is that similar terms 
could be inserted as well in eqs. \p{n42mod} with preserving $N=4$ 
supersymmetry. We will elaborate on these possibilities elsewhere.

It would be interesting to find the Lax operator (if existing) 
and the hamiltonian formulation for the above system, including 
the second hamiltonian structure superalgebra. We suspect that 
this system (or some its modification) could bear a tight relation 
to the super KdV hierarchy with the ``large'' $N=4$, 
$SO(4)\times U(1)$ superconformal algebra as the second hamiltonian 
structure. Indeed, inspecting the component contents of the relevant 
superfield set $J, \Phi, \bar{\Phi}, F_i, \bar{F}_i,\;i=1,2$, 
we find four dimension  $1/2$ and four dimension 
$3/2$ fermionic fields, as well as seven dimension 
$1$ and one dimension $2$ bosonic fields. This is just the 
currents contents of the "large" $N=4$ superconformal 
algebra \cite{{A},{belg}}. 
\vspace{0.3cm}

\noindent{\bf 5. Conclusions.}
In this Letter we constructed new $N=2$ supersymmetric integrable 
systems by junction of the pseudo-differential superspace Lax operators 
for $a=4$, $N=2$ KdV and multi-component $N=2$ 
NLS hierarchies. As a by-product we obtained Lax operator for 
$N=4$, $SU(2)$ super KdV system and thus proved the integrability 
of the latter. A similar extension of the $\alpha = -1/2$, $N=2$ 
super Boussinesq hierarchy was found. 

An intriguing characteristic feature of the proposed construction is the
possibility to extend some particular Lax operator by M additional
$N=2$ chiral and anti-chiral superfields. We are still not aware  
of the general recipe of how to construct such extensions, only 
two above examples have been explicitly worked out so far. 
Now it is under investigation whether the remaining two $N=2$ KdV 
and Boussinesq hierarchies (the $a=-2, 1$ KdV and $\alpha = -2, 5/2$ 
Boussinesq ones) can be extended in a similar way. 

It seems also very interesting to study in more detail the  
$N=4$ supersymmetric extension of the second flow of $N=4$ KdV hierarchy 
and to check its possible integrability. There remains a problem of 
putting this system, as well as the above Lax representation for 
$N=4$ KdV, into a manifestly $N=4$ supersymmetric form (e.g., in the 
framework of $1D$ harmonic superspace).
\vspace{0.3cm}

\noindent{\bf Acknowledgement.} We are indebted to F.~Delduc, L.~Gallot 
and A.~Sorin for valuable discussions. 
This work was supported by the grant of 
the Russian Foundation for Basic Research RFBR 96-02-17634, by INTAS 
grant INTAS-94-2317 and by a grant of the Dutch NWO Organization.

\end{document}